\documentclass[useAMS,usenatbib,letterpaper]{mn2e}
\usepackage{graphicx}
\usepackage{amsmath}
\usepackage{amssymb}

\title[Why and when is internally-driven AGN feedback energetically favoured?]{Why and when is internally-driven AGN feedback energetically favoured?}

\author[E.C.D. Pope] {Edward C.D. Pope$^{1}$\thanks{E-mail:ecdpope@uvic.ca}\\$^{1}$Department of
  Physics \& Astronomy, University of Victoria, Victoria, BC, V8P 1A1,
  Canada\\}

\begin{document}

\pagerange{\pageref{firstpage}--\pageref{lastpage} \pubyear{2011}}

\maketitle

\label{firstpage}

\begin{abstract}

AGN outflows are the heat given up when gas in a galaxy evolves towards thermodynamic equilibrium. Indeed, while AGN feedback regulates the 
growth of massive galaxies, its origins can be understood as the spontaneous thermodynamic process which ensures that the (Gibbs) free energy of the 
system always decreases, enabling the galaxy to reach a more energetically favourable state. In particular, it is shown that feedback heating processes will be favoured whenever the hot 
atmosphere of a galaxy would effectively gain energy as a result of cooling. For example, as the hot atmosphere of a galaxy cools and contracts, the work done by gravity will 
be thermalised, with a fraction of the gas also being captured by stars and the supermassive black hole at the centre of the galaxy. If this gain of energy exceeds the loss of 
energy that occurs when cooling gas drops out of the atmosphere, the Gibbs free energy of the system would increase overall. Since this is energetically unfavourable, feedback 
heating is initiated which acts to reduce the net cooling rate of the atmosphere, thereby preventing any build-up of energy. The Gibbs free energy can also decrease in the absence 
of feedback heating, but only if the loss of energy due to mass dropping out of the atmosphere exceeds the gains of energy described above. Therefore, to ensure that the Gibbs free 
energy always decreases, a galaxy will necessarily flip between these two states, experiencing episodes of heating and cooling. Due to the close long-term balance between heating 
and cooling, the gas in a galaxy will evolve quasi-statically towards thermodynamic equilibrium, which has the observable appearance of galaxy growth being regulated by AGN feedback. 
The same mechanism also provides an explanation for why strong AGN feedback occurs more frequently in cool-core galaxy clusters than in non cool-core clusters.

\end{abstract}


\section{Introduction}

The cooling time of X-ray emitting gas near the centres of many
massive galaxies and galaxy clusters is much shorter than the Hubble
time. In the absence of heat sources, significant quantities of the
gas would cool and form stars. However, X-ray spectroscopy has shown
that the rate at which gas cools to low temperatures is much
lower than first expected \citep[e.g.][]{peterson01, tamura01, xu02, sak, peterson03, kaastra04, peterson06}
suggesting that the gas is somehow being reheated.

A wealth of both observational and theoretical evidence suggest 
that energy input by a central Active Galactic Nucleus (AGN) is 
responsible for reheating the gas. In particular, powerful outflows 
produced by AGN interact strongly with their environment, inflating 
lobes of radio emission in the hot gas that permeates massive galaxies 
and clusters of galaxies \citep[e.g.][]{birzan,
  best05,dunn05,rafferty06,best08,shab08}. In addition, recent
observational studies of Brightest Cluster Galaxies (BCGs) suggest
that this radio AGN activity is related to the thermal state of its
environment. Systems with short radiative cooling times, or a low
central entropy, are more likely to exhibit active star formation,
optical line-emission and outflow-producing AGN
\citep[e.g.][]{burns,crawf,cav08,mittal,raff08}. 

Implementing and extending the ideas of early work \citep[e.g.][]{bintab,tucker,cfq,sr}, 
theoretical studies have drawn attention to the potential, wide-ranging 
impact of AGN feedback \citep[e.g.][]{andy99,nature, omma04, ka05, sijacki06, dimatteo08,booth09, soker9, mccarthy2010, gaspari2}. 
Notably, semi-analytic models of galaxy formation have demonstrated that, in principle, AGN heating can both reheat cooling 
flows and explain the exponential cutoff at the bright end of the galaxy
luminosity function \citep[e.g.][]{benson,croton05,bower06}, see also
\cite[][]{short}. In addition, AGN heating has been shown to be
crucial in shaping the X-ray luminosity-temperature relation of
massive galaxies \citep[e.g.][]{puchwein,bower08,pope09}. As such, AGN 
feedback has become a ubiquitous feature of galaxy evolution models, 
underpinning much of our understanding of why massive galaxies are they 
way they are.

However, the accuracy of galaxy evolution models is limited by their spatial 
resolution and our incomplete understanding of important physical 
processes -- particularly those that occur on sub-grid spatial scales, 
such as the the microphysics of the gas, and black hole accretion. As a result, it remains 
difficult to implement feedback in a fully self-consistent manner, hindering our understanding 
of galaxy evolution. For this reason, it is vitally important to better understand 
the fundamental processes that drive and trigger feedback. 

As an example of an AGN-triggering process that is not currently captured in cosmological models of galaxy formation, 
\cite{popeetal11} demonstrated that gas in the centre of a massive galaxy 
can become gravitationally unstable on sub-kiloparsec scales, thereby providing the AGN 
with an abundant source of fuel \citep[see][for different examples]{soker06, king09}. The characteristic duty cycles AGN activity
predicted by \cite{popeetal11}  are in agreement with observations \citep[e.g.][]{best05, best07, shab08}, suggesting the 
model does describe some important features of the AGN feedback phenomenon. As a result, this knowledge 
can be used to develop an improved prescription for AGN activity in a numerical model 
that does not explicitly account for self-gravity on sub-kiloparsec scales.

If correct, \cite{popeetal11} goes some way to explaining an important physical process that triggers AGN 
feedback in a hot atmosphere, but it does not explain why the system should evolve in such a way: why does 
AGN feedback occur at all? 

To re-phrase the problem: assuming that a galaxy can be considered to be a thermodynamic system, 
it should naturally evolve to minimise its free energy \citep[e.g.][]{lynden}\footnote{More specifically, a galaxy is likely to be an open 
thermodynamic system since it can exchange mass and energy with its environment. In contrast, a closed thermodynamic system is one 
that only exchanges energy with its environment, while an isolated system shares neither mass nor energy with its environment.} 
\citep[see also][for a discussion of galaxy cluster thermodynamics without feedback heating]{nusser09}. With this in mind, it is perhaps 
surprising that AGN feedback occurs at all since, by re-heating the galaxy's atmosphere, it appears to be obstructing the 
galaxy's tendency to evolve towards a minimum free energy configuration. 

At this point, it is necessary to distinguish between AGN heating which is driven by external influences 
such as galaxy mergers, and AGN heating that is driven internally by various properties of the host galaxy such as the 
X-ray luminosity, temperature and mass of its gaseous atmosphere. The latter constitutes AGN feedback in the truest 
sense of the definition and will be the focus of this paper. The distinction is relevant because merger events push the system 
further \emph{away} from thermodynamic equilibrium. In contrast, internally-driven processes occur spontaneously 
as the system evolves \emph{towards} thermodynamic equilibrium by reducing its free energy \citep[e.g.][]{fermi, statphys}. However, it is 
important to mention that merger-driven AGN heating may also play a significant role in galaxy evolution, with recent work suggesting that 
the effect may be more conspicuous for galaxies in group and field environments than in clusters \citep[see][for a recent example]{kaviraj}.

As indicated above, the progression towards thermodynamic equilibrium provides an informative framework with which to describe the 
characteristic features of gas evolution in a galaxy. Indeed, it is important to note that while the atmosphere of a galaxy may be close to hydrostatic 
equilibrium it is far from thermodynamic equilibrium either with its environment or internally. This is because the atmosphere is much hotter than the 
Cosmic Microwave Background, and it contains stars and a supermassive black hole which have a range of temperatures \citep[e.g.][]{hawk}. 
Therefore, according to the second law of thermodynamics, there will be large fluxes of heat from the hot atmosphere to its environment, and to the stars and the 
supermassive black hole. In particular, the very low temperature of a supermassive black hole means that any accretion of mass and energy from its environment 
results in a very large entropy increase \citep[e.g.][]{bek, bard, hawk} and is, thus, highly probable. However, the power output and subsequent heating 
that arises when the black hole grows prevent the accretion rate from rising uncontrollably. The competing influences of energy and entropy are, most generally, 
quantified by the free energy of the system \citep[e.g.][]{fermi, statphys}. Consequently, the free energy can be considered to be a thermodynamic potential. To date, 
this approach has proved to be useful in describing stable states of gravitating gas spheres \citep[e.g.][]{lynden, stahler, tomisaka, mckee99, bludman}. In this article, 
we find the Gibbs free energy to be the most suitable form of thermodynamic potential with which to describe and explain the driving force behind the feedback processes 
thought to be key in governing the evolution of the galaxy. 

Formally, a thermodynamic system will achieve a state of thermodynamic equilibrium when its free energy reaches a minimum \cite[e.g.][]{fermi}. As a result, for systems 
that have not yet reached equilibrium, processes that reduce the free energy will be favoured and will occur spontaneously \citep[e.g.][]{fermi, lynden}. 
In this formalism, AGN and stellar feedback are the spontaneous processes which ensure that the free energy of the system 
always decreases, so that the system tends towards thermodynamic equilibrium. Therefore, it is shown that AGN feedback 
\emph{is} energetically favourable. This is primarily because feedback heating prevents the effective build up of energy that occurs as hot gas from the galaxy's 
atmosphere cools and flows towards the centre of the galaxy. As this happens, the stored gravitational potential energy of the atmosphere is converted into heat, while 
significant quantities of the cooling gas will be gravitationally captured by stars and the supermassive black hole at the centre of the galaxy. Gas that is gravitationally 
captured by a compact object will be bound to it with an energy that is a non-negligible fraction of its rest mass energy. Since this is much greater 
than its binding energy to the host galaxy, the gas can effectively gain energy as it cools\footnote{This phenomenon is similar to the negative heat 
capacity of galaxy clusters discussed by \cite{nusser09}. However, it differs because feedback heating intervenes to prevent the heat capacity 
becoming negative.}. Finally, as the gas falls onto the compact object, the binding energy will be radiated as heat, reducing the net cooling rate 
of the atmosphere. Consequently, feedback prevents the build up of energy that can occur when gas cools in the presence of compact objects, 
and ensures that the galaxy evolves slowly towards a minimum free energy configuration.

The article is organised as follows: in section 2 we derive a general thermodynamic
galaxy model, in section 3 we consider the broader implications of the model, and 
summarise the findings in section 4.

\section{Model}

While this model mainly focuses on explaining the origin of feedback heating in massive elliptical galaxies, 
the principles also hold for stellar feedback, and AGN feedback in galaxy clusters. 

To ensure the study is applicable to the widest range of environments we start with a thermodynamic 
description of a hot galaxy atmosphere (with an embedded cool phase) that flows within the galaxy's darkmatter gravitational potential. 
The gravitational field exerted by the darkmatter is comparatively smooth and extends far beyond the visible extent of the galaxy. It is this 
field that governs the large-scale properties of the hot atmosphere. Differences in the global heating and cooling rates drive the transfer of 
mass between the hot and cool phases. Gas belonging to the cool phase is assumed to be strongly influenced by the 
presence of stars and the black hole since it can be more easily accreted than hotter gas, due to its lower thermal velocity. For this reason, the 
approach is extended to consider a generalised galaxy model which describes the 3 main gravitational influences on the gas: 
i) the darkmatter; ii) the stars; iii) the supermassive black hole. The system then favours heating and cooling process which cause the galaxy's gas 
to move between the gravitational components, re-distributing mass and energy, in such a way that it evolves towards a thermodynamic 
equilibrium state. 

\subsection{Total energy of a hot gaseous atmosphere}

The hot gaseous atmosphere that permeates a massive galaxy is thermally ionised, and cools through the emission of X-rays. 
As the gas radiates energy it loses pressure support so that the weight of the overlaying gas causes an inflow. Nevertheless, 
the gas is generally assumed to be close to hydrostatic equilibrium because the cooling time is much longer than the free-fall 
time \citep[e.g. see][ for a review]{mcnuls}. Additional deviations from the hydrostatic limit are caused by bulk flows and turbulent eddies driven 
by merger events and large-scale outflows from the galaxy \citep[e.g.][]{ricker01,ds08,scanna,stern,dong,pelligrini}. 
As such, the evolution of a hot atmosphere can be described in terms of its kinetic, thermal and gravitational potential energy. 

For the purposes of this study, it is necessary to define a region of interest within the galaxy atmosphere. The standard choice 
is the volume enclosed by the cooling radius -- the radius at which the gas cooling time equals (a fraction of) the Hubble time. The reason for this choice would be that the volume enclosed 
by the cooling radius provides an estimate of the gas mass that can have cooled within the lifetime of the Universe. The cooling radius, $r_{\rm cool}$,  often implicitly enters calculations 
which relate the bolometric X-ray luminosity of the hot gas, $L$, within $r_{\rm cool}$, its average temperature, $T$, within the same volume, and the so-called classical mass flow rate, 
$\dot{m}_{\rm clas}$. The magnitude of this flow rate is given by \citep[e.g.][]{fab94} 
\begin{equation}\label{eq:m__clas}
L = \frac{\gamma}{(\gamma-1)}\frac{k_{\rm b}T}{\langle m \rangle}\dot{m}_{\rm clas},
\end{equation}
where $\gamma$ is the adiabatic index of the hot gas, $k_{\rm b}$ is Boltzmann's constant and $\langle m \rangle$ is the mean mass per particle. According to the derivation of equation (\ref{eq:m__clas}) 
from the fluid energy equation, $\dot{m}_{\rm clas}$ is the mass flow rate due to the contraction of the hot gas of temperature $T$ as it radiates energy at a rate $L$. As such, $\dot{m}_{\rm clas}$ is the rate at which mass flows through $r_{\rm cool}$ due to the contraction of the hot gas within $r_{\rm cool}$. More generally, the mass flow rate across $r_{\rm cool}$ is determined by the difference between the global heating rate, $H$, and the cooling rate, $L$, such that $H-L \propto \dot{m}_{\rm tot}$.  In the limit that the change in gravitational potential energy is small by comparison, the mass flow rate of the hot atmosphere takes the following form \citep[e.g.][]{sarazin} 
\begin{equation}\label{eq:10}
H-L \approx \frac{1}{(\gamma -1)}\dot{m}_{\rm tot}c_{\rm s}^{2}, 
\end{equation}
where $c_{\rm s}^{2} = \gamma k_{\rm b}T/\langle m \rangle$ is the sound speed of gas in the atmosphere. Thus, if $H < L$, the atmosphere cools overall, and mass will flow inwards. 
If $H > L$, the gas is heated overall, and mass will flow outwards. It is important to note that $H$ is the rate at which the injected energy is dissipated as heat; as such, it represents a fraction, $\eta_{\rm c}$, 
of the total energy injected to the system. For example, an AGN accreting material at a rate less than $\sim$3\% of the Eddington limit is likely to be radiatively inefficient so that the majority 
of the power output is in the form of kinetic outflows that are thought to couple strongly to the ambient gas, i.e. $\eta_{\rm c} \sim 1$. Conversely, accretion above this critical rate is radiatively 
efficient, meaning that the power output is predominantly in the form of photons. In the radiatively efficient mode, it has been argued that only $\eta_{\rm c} \sim 5\%$ of the accretion power is available to 
heat the surrounding gas \citep[e.g.][]{sijacki06,king09}. The coupling parameter $\eta_{\rm c}$ can also be used to account for the fraction of energy dissipated within the region of interest for outflows that 
extend beyond its outer boundary.

While the cooling radius provides a reasonable choice for an outer boundary under many circumstances, the changing mass within $r_{\rm cool}$ and the unknown variation of the external pressure 
at $r_{\rm cool}$ make calculations of the work done on the region of interest much more complex than they need to be. For this reason, the calculations presented in this investigation make use of a more 
appropriate outer boundary condition that: i) simplifies the calculation of work done on the region of interest; ii) ensures there is no mass flow across this boundary, as shown below, meaning that the system 
is formally closed. In particular, we choose an outer boundary which is defined by a virtual surface at fixed pressure ($P_{\rm ext}$) that moves inwards and outwards depending if the gas within is cooled or 
heated overall. To prove that there should be no mass flow across an outer boundary of fixed pressure, for quasi-static changes, we present the following argument. For a region of gas bounded by an 
external pressure $P_{\rm ext}$, heated at a rate $H$ and cooled at a rate $L$, the rate of change of volume is given by $H-L = P_{\rm ext} \dot{V} \gamma/(\gamma-1)$. Using the ideal gas law, the 
pressure is written $P_{\rm ext} = \rho k_{\rm b}T/\langle m \rangle$, where $\rho$ and $T$ are the density and temperature at the boundary. Here, we assume that the temperature throughout the atmosphere is constant, so that the temperature at the outer boundary is equal to the average temperature. 
Then, assuming that the region is spherical, the rate of volume change is  $\dot{V} = 4\pi r^{2} \dot{r}$, with $\dot{r}$ being the rate of change of radius at which the external pressure is $P_{\rm ext}$. 
Substituting for $P_{\rm ext}$ and $\dot{V}$ gives
\begin{equation} \label{eq:equiv}
H-L = 4\pi r^{2} \rho  \dot{r} \frac{\gamma}{(\gamma-1)}\frac{k_{\rm b}T}{\langle m \rangle} = \frac{1}{(\gamma -1)}\dot{m} c_{\rm s}^{2}.
\end{equation}
Thus, for an appropriate choice of $P_{\rm ext}$, the gas volume of interest will change radius at approximately the same velocity as the local mass flow given by 
equation (\ref{eq:10}), meaning that $\dot{m}  \approx \dot{m}_{\rm tot}$. Therefore, under ideal conditions, there should be no mass flow across the boundary defined by constant pressure, \emph{but} it is important to remember that, due to differences between the global heating and cooling rates, there is still mass flowing within the region, i.e. relative to a fixed radius. In particular, the mass flow rate relative to a fixed cooling radius has a magnitude of $\approx \dot{m}_{\rm tot}$. For this reason, and to ensure the region is large enough that feedback energy is dissipated within it \citep[e.g.][]{h06, pope09, fabj}, $P_{\rm ext}$ is taken to be the pressure at the 
initial cooling radius. Finally, the region of interest is assumed to be spherical and centred on the supermassive black hole at the centre of the galaxy. Under these assumptions, the total energy 
of the gaseous atmosphere within the region of interest, $\Omega$, can be written as
\begin{equation} \label{eq:1}
E(\Omega) = K(\Omega) + \Theta(\Omega) + \Psi(\Omega),
\end{equation}
where $K$ is the bulk kinetic energy, $\Theta = U + P_{\rm ext}V$ is the enthalpy of the gas volume of interest, with $U$ 
being the internal energy and $V$ being the volume of $\Omega$, and $\Psi$ is the gravitational potential energy. 

As the gas within $\Omega$ cools, the gas will lose gravitational potential energy while simultaneously having compression work done on it. Initially, 
both the gravitational and compression work will increase the kinetic energy of the gas. However, as long as the relaxation time of 
the gas is short compared to the work timescale, the kinetic energy will be thermalised within $\Omega$. In this limit, the kinetic energy of the 
system will remain constant, the system will evolve quasi-statically, and the compression can be considered to be reversible. The 
relationship between the kinetic, thermal and potential energies can be derived by differentiating equation 
(\ref{eq:1}) with respect to time
\begin{equation}\label{eq:2}
\dot{E} = \dot{K} + \dot{\Theta} + \dot{\Psi},
\end{equation}
where we have omitted $\Omega$ for brevity.

According to the argument above, the rate of change of kinetic energy is equal to the total rate at which work is done on the system, less 
the rate at which the kinetic energy is thermalised, $\dot{D}$. The total rate at which work is done on the system is sum of the rate at which 
gravity does work ($-\dot{\Psi}$) and the rate of compressional work done on the system ($\dot{W}$), such that 
$\dot{K} = \dot{W} - \dot{\Psi} - \dot{D}$. Then, in the limit that the work is performed reversibly within $\Omega$, we can write 
$\dot{D} = \dot{W} - \dot{\Psi}$. 

By definition, the rate of change of enthalpy is given by the sum of the heat fluxes, $\dot{Q}$, plus the kinetic energy dissipation rate, 
less the rate of compression work; $\dot{\Theta} = \dot{Q} + \dot{D} -\dot{W}$. Thus, in the limit of reversibility, we have 
$\dot{\Theta} = \dot{Q}  - \dot{\Psi}$, which is the first law of thermodynamics for a system in which the change of gravitational potential 
energy is important. Finally, the rate of change of total energy is $\dot{E} = \dot{Q} + \dot{W}$.

Under the assumption that the hot gaseous atmosphere evolves quasi-statically, it is appropriate to use further thermodynamic arguments
to describe the direction of processes driving this evolution. These arguments are outlined and explored below.

\subsection{The Gibbs Free Energy of a galaxy atmosphere}

A system will be in stable thermodynamic equilibrium when its free energy is at a minimum \citep[e.g.][]{fermi}. As a result, processes that reduce 
the free energy will be favoured and will occur spontaneously \citep[e.g.][]{lynden}. In contrast, processes that increase the free energy are energetically 
unfavourable and will be improbable. 

The appropriate form of the free energy for a system at constant pressure $P_{\rm ext}$ and temperature $T$, is the Gibbs free energy -- henceforth 
referred to as GFE -- which is defined as
\begin{equation}\label{eq:3}
G = \Theta - TS
\end{equation}
where $\Theta$ is the enthalpy as defined previously, $T$ is the system temperature and $S$ is the system entropy. 

Equation (\ref{eq:3}) should be viewed as a thermodynamic potential that quantifies the relative influences of enthalpy and 
entropy. Thus, for an isolated system (with constant $U$, $P_{\rm ext}$ and $V$), the GFE 
is minimised by maximising the entropy. In contrast, for a closed system that exchanges energy with its environment, 
the GFE will be minimised by decreasing the enthalpy and increasing the entropy \citep[e.g.][]{lynden}. 

Following the definition of the GFE, a system will be in equilibrium when $G$ is a minimum; this occurs when 
the change in GFE is zero, i.e. $\Delta G = 0$ \citep[e.g.][]{fermi,lynden}. Using equation (\ref{eq:3}), the change in the GFE is written
\begin{equation}\label{eq:4}
\Delta G = \Delta \Theta - T\Delta S - S \Delta T.
\end{equation} 
To generate a more useful form of the differential GFE, we substitute for $\Delta \Theta$ and $T\Delta S$, and assume that the 
temperature of the atmosphere remains approximately constant so that $S \Delta T = 0$.  This assumption is justified because 
the cooling time of the gaseous atmosphere is generally much longer than the gravitational free-fall time. Therefore, whether the 
gas is being heated (expanding) or cooled (contracting), the gas will flow sufficiently slowly that it can be considered to be close 
to hydrostatic equilibrium, i.e. its sound speed must comparable to the velocity dispersion of the underlying gravitational potential. Since
the gravitational potential is taken to be static, the gas temperature must also be approximately constant.

From its definition, the change of enthalpy is written $\Delta \Theta = \Delta U + P_{\rm ext}\Delta V + V\Delta P_{\rm ext}$. Since the 
outer boundary of $\Omega$ is defined to be movable so as to maintain constant pressure ($\Delta P_{\rm ext} = 0$) the 
differential enthalpy becomes $\Delta \Theta = \Delta U + P_{\rm ext}\Delta V$. Furthermore, the second law of thermodynamics for 
the reversible transfer of heat, $\Delta Q$, from a system at temperature, $T$, to its environment, is written
\begin{equation}\label{eq:5}
\Delta Q = T \Delta S
\end{equation}
where $\Delta S$ is the entropy change of the system. Accordingly, equation (\ref{eq:4}) becomes 
\begin{equation}\label{eq:6}
\Delta G = \Delta U + P_{\rm ext}\Delta V - \Delta Q.
\end{equation}
To further simplify equation (\ref{eq:6}), we substitute for $\Delta U$ using the first law of thermodynamics for a system in which 
any work done by gravity is immediately thermalised. In addition, even though there is no mass flow across 
the outer boundary of the system, we still include a mass transfer term ($\mu \Delta N$) to account for cooling gas that drops out 
of the hot atmosphere to join the cool phase described previously. At this point is it is important to note that the cool phase is assumed 
to act as a reservoir that will return mass to the hot phase whenever the global heating rate exceeds the cooling rate. 

The general form of the first law of thermodynamics can be written
 \begin{equation}\label{eq:7}
\Delta U =  \Delta Q  - P_{\rm ext}\Delta V - \Delta \Psi +  \mu \Delta N,
\end{equation}
where $\mu$ is the energy per transferred particle, and $\Delta N$ is the change of number of particles in the system, due 
to mass dropping out of the atmosphere, or being added to it. It is important to note again that $-\Delta \Psi$ is the change of internal energy that 
arises due to the gravitational work being thermalised. By substituting equation (\ref{eq:7}) in to equation (\ref{eq:6}), the change in GFE of the hot 
atmosphere can be written
\begin{equation}\label{eq:8}
\Delta G = - \Delta \Psi + \mu_{\rm atmos} \Delta N.
\end{equation}
Therefore, as the gas cools and loses gravitational potential energy ($\Delta \Psi < 0$), the GFE 
will increase unless mass drops out of the atmosphere such that $|\mu_{\rm atmos} \Delta N| > |\Delta \Psi|$. 

To make equation (\ref{eq:8}) directly applicable to a galaxy atmosphere, we express the gravitational and mass loss
terms as functions of the global heating and cooling rates. In the limit that self-gravity of the atmosphere can be 
ignored, the change of gravitational potential energy due to mass flow within the galaxy's darkmatter gravitational field, 
during an interval $\Delta t$, can be written as
\begin{equation}\label{eq:9}
\Delta\Psi \approx \alpha \dot{m}_{\rm tot} \sigma_{\rm DM}^{2}\Delta t, 
\end{equation}
where $\dot{m}_{\rm tot}$ is the mass flow rate within the hot atmosphere, $\sigma_{\rm DM}$ is 
the velocity dispersion of the darkmatter potential, and $\alpha$ is a numerical constant the quantifies the gravitational potential 
energy of the gas within $\Omega$ as a fraction of $\sigma_{\rm DM}^{2}$. The total mass flow rate is determined by the difference 
between the global heating and cooling rates in the atmosphere, as shown in equations (\ref{eq:10}) and (\ref{eq:equiv}).
The energy flux due to mass transfer between the hot atmosphere and the embedded cool phase is derived using the following considerations. 
The net particle flux into or out of the cool phase in a time interval $\Delta t$ will be $\Delta N = \beta \dot{m}_{\rm tot} \Delta t/\langle m \rangle$, where $\beta \dot{m}_{\rm tot}$ 
is the mass flux at which hot gas cools and drops out of the atmosphere. For convenience, we have expressed this mass dropout rate as a fraction of the bulk mass flow rate, $\dot{m}_{\rm tot}$, which occurs whenever the global cooling and heating rates are not equal, as seen in equation (\ref{eq:10}). A consequence of this argument is that a mass fraction $1-\beta$ of the hot atmosphere does not cool significantly, despite being part of the bulk flow. Gas which drops out of the flow is assumed to join the cool phase. As mentioned above, this acts as a reservoir that can return material back to the hot phase when the global heating rate exceeds the cooling rate. For simplicity, we assume that the mass returned from the cool reservoir makes up the same fraction, $\beta$, of the bulk mass flow rate, $\dot{m}_{\rm tot}$. 
Consequently, the expression $\Delta N = \beta \dot{m}_{\rm tot} \Delta t/\langle m \rangle$ applies whether the gas is being cooled or heated.

The particles are assumed to be removed from and returned to the flow at the temperature of the hot phase, meaning that $\mu_{\rm atmos}  = \gamma k_{\rm b}T/(\gamma-1)$. 
For completeness, a net heating rate of $\beta(H-L)$ is required to re-heat cool phase material up to the temperature of the hot atmosphere at a mass flow rate of $\beta \dot{m}_{\rm tot}$. 
This represents the dominant input of energy required to return cool phase material back to the hot phase flow, i.e. since the hot phase flow is highly subsonic, the bulk kinetic energy of the 
heated material is negligible in comparison to its thermal energy. Then, accounting the work done by gravity and mass transfer, the change in GFE can be written
\begin{equation}\label{eq:11}
\Delta G = (H-L)\beta \bigg[1 - \frac{\alpha}{\beta}(\gamma-1)\bigg(\frac{\sigma_{\rm DM}}{c_{\rm s}}\bigg)^{2}\bigg]\Delta t.
\end{equation}
Accordingly, we see that the system will favour strong heating (i.e. $H > L$) if $(\alpha/\beta)(\gamma-1)\sigma_{\rm DM}^{2} > c_{\rm s}^{2}$, i.e. 
\emph{if the gas temperature is cool compared to the effective dynamical temperature of the gravitational potential}. In this case, 
the increase of internal energy resulting from the work done by gravity exceeds the energy that is lost by mass dropping out of the 
atmosphere. In contrast, the system will favour cooling (i.e. $L > H$) if $(\alpha/\beta)(\gamma-1)\sigma_{\rm DM}^{2} > c_{\rm s}^{2}$, 
i.e. \emph{if the gas temperature is warmer than the effective dynamical temperature of the gravitational potential}. In this case, 
increase of gas internal energy resulting from the work done by gravity is less than the energy that is lost by mass dropping out of 
the atmosphere.

Assuming that the heating is provided by an AGN at the centre of the galaxy then, provided the black hole is massive enough, strong heating (i.e. $H > L$) 
will occur at a lower accretion rate for radiatively inefficient accretion than for radiatively efficient accretion. As described earlier, this is because the kinetic outflows couple 
much more strongly to the ambient gas than the photons produced by radiatively efficient accretion. Since the black hole accretion rate is likely 
to increase from a very low value during a cooling phase (i.e. $L > H$), this suggests that the first incidence of strong heating will occur at a low Eddington fraction, meaning that the 
AGN power output will be in the form of kinetic outflows. While $H > L$, the black hole accretion rate will be prevented from increasing significantly thereby ensuring that the AGN 
persists in the radiatively inefficient state. If this argument is correct, it suggests that internally-driven AGN outbursts should be predominantly in the form of kinetic outflows. However, in the case that 
radiatively inefficient accretion is insufficient to provide strong heating, the accretion rate will continue to increase, causing the AGN to enter the radiatively inefficient accretion state. Since the power output 
from the AGN in this state is less effective at heating the ambient gas, the accretion rate will continue to increase. As a result, the black hole will grow rapidly until it is massive enough for 
radiatively inefficient power output to provide strong heating, as described in \cite{chur06}.

Importantly, differences between the global heating and cooling rates will modify the value of $(\alpha/\beta)(\gamma-1)(\sigma_{\rm DM}/c_{\rm s})^{2}$. 
This is because $\alpha$, being a function of the gas density within $\Omega$, $c_{\rm s}$ and possibly will depend on the historical 
balance between heating and cooling, as well as the underlying gravitational potential. Therefore, we should expect 
$(\alpha/\beta)(\gamma-1)(\sigma_{\rm DM}/c_{\rm s})^{2}$ to decrease when the gas is heated overall, and to increase when the gas is cooled
overall. Consequently, the energetics of the atmosphere favour a heating episode while $(\alpha/\beta)(\gamma-1)\sigma_{\rm DM}^{2} > c_{\rm s}^{2}$. 
However, when the gas has been heated sufficiently that $(\alpha/\beta)(\gamma-1)\sigma_{\rm DM}^{2} > c_{\rm s}^{2}$, the energetics will favour
a period during which cooling is favoured. This state will persist until $(\alpha/\beta)(\gamma-1)\sigma_{\rm DM}^{2} < c_{\rm s}^{2}$, at which point
another heating episode will be favoured.

Equation (\ref{eq:11}) suggests a general solution to the characteristic differences of AGN activity observed in cool-core (CC) and non 
cool-core (NCC) galaxy clusters \citep[e.g.][]{burns,mittal}. According to equation (\ref{eq:11}), the lower fraction of AGN outbursts in NCC clusters 
compared to similar CC clusters can be explained by differences in the ratio of the dynamical temperature of the gravitational 
potential and the thermal temperature of the atmosphere. Therefore, we should expect that NCC clusters spend more time with 
$(\alpha/\beta)(\gamma-1)(\sigma_{\rm DM}/c_{\rm s})^{2} < 1$, while CC clusters spend more time with $(\alpha/\beta)(\gamma-1)(\sigma_{\rm DM}/c_{\rm s})^{2} > 1$. 
These differences could arise due to the formation histories of galaxy clusters; for example, NCC clusters may have experienced 
more energetic mergers than CC clusters, thereby disrupting the cooling flow and increasing the thermal energy of the gas relative 
to the dynamical temperature of the gravitational potential \citep[e.g.][]{mcglynn,burns97,gomez,poole08}. The effect of thermal conduction 
\citep[e.g.][]{guo} and differences in metal content \citep[e.g.][]{dubois} may also play a role in this behaviour.

However, while equation (\ref{eq:11}) describes the change of GFE for the hot atmosphere, it does not fully represent the energetics of
gas in the galaxy. In particular, due to its lower thermal velocity, the cool phase material can be gravitationally captured by stars and the 
supermassive black hole. As described in the introduction, the gravitational binding energy 
of gas captured by a compact object is a non-negligible fraction of its rest mass energy. Since this is much greater 
than its binding energy to the darkmatter gravitational field, some of the cooling gas can effectively gain energy. Then, as the gas 
falls onto the compact object, the binding energy will be radiated as heat, thereby reducing the net cooling rate of the atmosphere. 
It is important to note here that the total potential energy does not increase, it only appears to because the rest mass energy of the gas 
becomes accessible in the presence of strong gravitational fields. 

To account for the influence of the stellar and black hole galaxy components, we return to the composite galaxy 
model which describes the 3 main gravitational influences on the gas: i) the darkmatter gravitational potential; ii) the stellar component; iii) the supermassive 
black hole. The galaxy's gas is assumed to move between each of these components as the system evolves towards a thermodynamic equilibrium state. 
Therefore, since these 3 galaxy components are closely coupled by mass and energy transfer, the overall system will tend towards 
equilibrium as long as the sum of the component free energies decreases.

\begin{figure*}
\centering
\includegraphics[width=10cm]{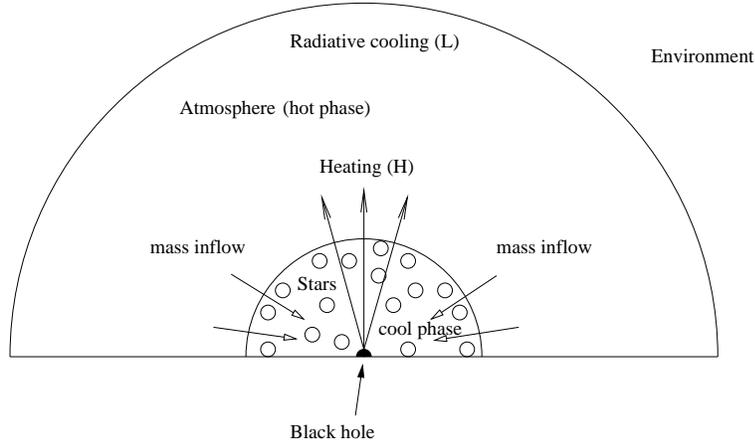}
\caption{Illustration of one hemisphere of the 3-component galaxy model consisting of the atmosphere, stars, and the black hole. As the atmosphere cools, mass will
flow inwards, with a fraction dropping out to join the cool phase (indicated by the clouds) which can be captured by the stars and the supermassive black hole. The boundary 
between the atmosphere and the environment is at a pressure $P_{\rm ext}$.}
\label{fig:gibbs_1}
\end{figure*}

\subsection{Gibbs free energy of the composite galaxy system}

In this extension of the arguments presented above, we make the approximation that the cool phase gas will be gravitationally captured by stars and the supermassive black hole.
Consequently, each of the 3 gravitational components will contain its own quantity of gas, as shown in figure \ref{fig:gibbs_1}. Thus there is a GFE for each gravitational component. 
Furthermore, due to cooling and heating processes, mass and energy will be re-distributed between the components, meaning that the component GFEs will change as the system 
evolves towards thermodynamic equilibrium. It is important to note here that the vast majority of the cool phase gas is assumed to be close to infinity with respect to the black hole 
and the stars. As a result, it takes a negligible quantity of energy to unbind the gas and re-distribute it among the different gravitational components. \emph{Then, since the gas 
components are closely coupled, the system will tend towards equilibrium if the sum of the changes of the component GFEs is less than zero, i.e. $\sum \Delta G< 0$, even if the GFE 
of any component increases}. 

The GFE of the stellar and black hole components each take the same form and consist only
of mass transfer terms which represent the change of energy as mass flows between the 
components. Accordingly, the change of GFE for gas that becomes bound to stellar gravitational fields is
\begin{equation}\label{eq:12}
\Delta G_{\rm *} = \mu_{\rm *} \Delta N_{\rm *}
\end{equation}
where $\mu_{\rm *}$ is the energy per particle of gas that is bound to the stars, and $\Delta N_{\rm *}$ is the
particle flux into the stellar component. 
The energy gain due to thermally cool gas being captured by stellar mass gravitational fields, during an interval $\Delta t$, can be written as
\begin{equation}\label{eq:13}
\mu_{\rm *} \Delta N_{\rm *} = \eta_{\rm c *}\eta_{*} \dot{m}_{*} c^{2}\Delta t, 
\end{equation}
where $\dot{m}_{\rm *}$ is the rate at which mass is captured by stellar mass gravitational fields, and $\eta_{*} c^{2}$ 
represents the average gravitational binding energy of gas to this component as a fraction of its rest mass energy, with 
$c$ being the speed of light, and $\eta_{\rm c *}$ is the fraction of energy released that could couple to the ambient gas.

The change of GFE for cool gas that becomes gravitationally bound to the supermassive black hole is
\begin{equation}\label{eq:14}
\Delta G_{\rm BH} = \mu_{\rm BH} \Delta N_{\rm BH},
\end{equation}
where $\mu_{\rm BH}$ is the energy per particle of gas that is bound to the black hole, and $\Delta N_{\rm BH}$ is the
particle flux into the back hole component during a time interval $\Delta t$. As above, the gain in energy of gas being captured by the 
black hole during the interval $\Delta t$, can be written as
\begin{equation}\label{eq:15}
\mu_{\rm BH} \Delta N_{\rm BH} = \eta_{\rm c}\eta_{\rm BH} \dot{m}_{\rm BH} c^{2}\Delta t, 
\end{equation}
where $\dot{m}_{\rm BH}$ is the rate at which mass is captured by the black hole gravitational field, and $\eta_{\rm BH} c^{2}$ 
represents the average gravitational binding energy of gas to this component as a fraction of its rest mass energy, and $\eta_{\rm c}$ 
is the fraction of energy released that could couple to the ambient gas.

In this model we assume that all of the mass entering the cool phase is captured either by the stars or the black hole, therefore it follows that 
$\beta \dot{m}_{\rm tot} + \dot{m}_{*} + \dot{m}_{\rm BH} = 0$. Then, if the mass joining the stellar component is a fraction $(1-f)$ of the 
mass entering the cool phase, we have $\dot{m}_{*} = -(1-f)\beta \dot{m}_{\rm tot}$; correspondingly, 
the mass captured by the black hole must be a fraction, $f$, of the mass entering the cool phase, $\dot{m}_{\rm BH} = -f\beta\dot{m}_{\rm tot}$. 

Combining the relations above means that the sum of the component GFEs is
\begin{eqnarray}\label{eq:16}
\sum \Delta G  = \bigg[-\alpha \dot{m}_{\rm tot} \sigma_{\rm DM}^{2} + \beta\dot{m}_{\rm tot} \frac{c_{\rm s}^{2}}{(\gamma-1)}\bigg] \Delta t  \\ \nonumber  + \bigg[-\eta_{\rm c *}\eta_{*} (1-f) \beta \dot{m}_{\rm tot} c^{2} - \eta_{\rm c}\eta_{\rm BH} f \beta \dot{m}_{\rm tot} c^{2} \bigg] \Delta t .
\end{eqnarray}
Using the definition of $\dot{m}_{\rm tot}$ from equations (\ref{eq:10}) and (\ref{eq:equiv}), the total change of the GFE is expressed in equations (\ref{eq:17} ) and (\ref{eq:18}). Specifically, 
these equations give the change in GFE for gas which is being heated at a total rate $H$, cooled at a rate $L$, accounting for the thermalisation of gravitational work 
and mass transfer to/from the atmosphere, \emph{and} the gravitational influence stars and the supermassive black hole,
\begin{equation}\label{eq:17}
\sum \Delta G = (H-L)\beta \bigg[1 - \bigg(\frac{v}{c_{\rm s}}\bigg)^{2}\bigg]\Delta t,
\end{equation}
where 
\begin{equation}\label{eq:18}
v^{2} \equiv (\gamma-1)\bigg[\frac{{\alpha}}{\beta}\sigma_{\rm DM}^{2} + (1-f)\eta_{\rm c *}\eta_{\rm *} c^{2} + f\eta_{\rm c}\eta_{\rm BH} c^{2}\bigg].
\end{equation}
Here, $v^{2}$ can be considered to be the energy per unit mass gained by the system due to the processes associated with gas cooling. Under the assumption that 
most internally-driven accretion events produce kinetic AGN outflows which dissipate their energy within the region of interest, the best choice for $\eta_{\rm c}$ is $\sim 1$. 
Furthermore, the value of $\eta_{\rm *} c^{2}$ in equation (\ref{eq:18}) is equivalent to the kinetic energy per unit mass of stellar winds or supernova ejecta used in semi-analytic galaxy evolution models 
\citep[e.g.][]{croton05,dave}. Comparison with such models suggests that $(1-f)\eta_{\rm *}$ will be much smaller than $f\eta_{\rm c}\eta_{\rm BH}$, provided $f$ is greater than $\sim 10^{-5}$, 
meaning that the value of $\eta_{\rm c *}$ is generally less important than $\eta_{\rm c}$.

\begin{figure*}
\centering
\includegraphics[width=10cm]{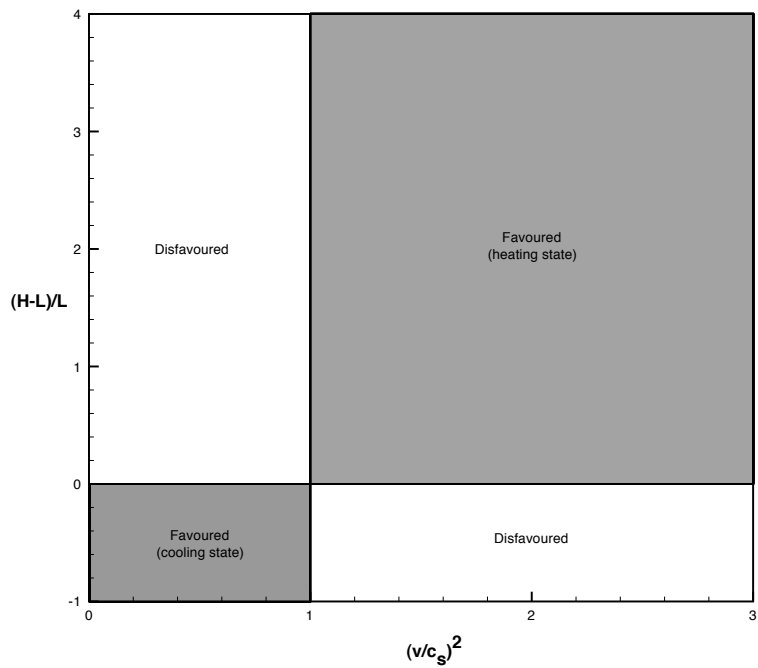}
\caption{Parameter space for which the Gibbs Free Energy decreases. Systems experiencing feedback 
heating will occupy the shaded region in the top right of the figure, where $H>L$, and $(v/c_{\rm s})^{2} > 1$.  In contrast, systems that are cooling will occupy the 
shaded region in bottom left of the figure, where $L>H$, and $(v/c_{\rm s})^{2} < 1$. A system will switch between these favoured states, depending on the ratio 
$(v/c_{\rm s})^{2}$ which will vary over time as the gas is heated and cooled. The $y$-axis ranges from -1, for $H=0$, to an arbitrary 
upper limit of +4, which implies $H = 5L$.  Similarly, the $x$-axis ranges from 0, for $v=0$, to an arbitrary upper limit of 
$(v/c_{\rm s})^{2} = 3$. The area of a region is not an indication of the probability of the system
occupying that state.}
\label{fig:gibbs}
\end{figure*}

Since the system favours changes that reduce the total GFE ($\sum \Delta G < 0$), heating will be favoured ($H>L$) if  $v >  c_{\rm s} $. 
That is, heating is favoured when the gain of energy that occurs when gas is captured by stars and the black hole $+$ the thermalisation of gravitational work, 
exceeds the loss of energy that arises when gas cools out of the hot atmosphere. The system can also favour cooling ($L > H$), but only if $v < c_{\rm s}$. 
That is, cooling will be favoured if the build up of energy in the presence of the stars and the supermassive black hole $+$ the thermalisation of gravitational work, 
is less than the loss that arises when gas cools out of the atmosphere. It is worth noting here that if $\eta_{\rm c} \sim 0.05$ the system is more likely, for a given value of $f$,
to be in a cooling state since $f\eta_{\rm c}\eta_{\rm BH}c^{2}$ will be smaller relative to $c_{\rm s}^{2}$. Figure \ref{fig:gibbs} shows the regions of parameter space which are 
favoured by systems that can be described using equation (\ref{eq:17}).

For the reasons outlined in the previous subsection, it is important to remember that differences between heating and cooling rates will modify the value of 
$(v/c_{\rm s})^{2}$ such that we expect $(v/c_{\rm s})^{2}$ to decrease when the gas is heated overall, and to increase when the gas is cooled overall. 
Therefore, the energetics of the system favour a heating episode while $v^{2} > c_{\rm s}^{2}$. However, when the gas has been heated sufficiently that 
$v^{2} > c_{\rm s}^{2}$, the energetics will favour a period during which cooling is favoured. Again, this will occur until $v^{2} < c_{\rm s}^{2}$, at which point 
another heating episode will be favoured. This means that the time-averaged heating rate will closely match the time-averaged cooling rate. Consequently, 
the system will evolve quasi-statically towards a thermodynamic equilibrium state. \emph{Quasi-static evolution has the 
observable appearance of feedback heating regulating the growth of its host galaxy}.

\section{Discussion}

Equation (\ref{eq:17}) is a general expression that describes the circumstances 
under which feedback heating will be favoured. However, although the formalism predicts that there should be periods of overall heating 
and cooling, it does not make any explicit predictions about the duration or frequency of these episodes. In order to solve this problem it 
would be necessary to derive and solve a differential equation for the time-evolution of $v^{2}$, as defined in equation (\ref{eq:18}). Any differential equation 
of this type would be highly uncertain. Indeed, it was for this reason that \cite{pope2011} used a different approach: to find a Lagrangian 
function that yielded periodic episodes of feedback heating that were in approximate agreement with the observations of 
\cite{best05, best07, shab08}. Notably, the Lagrangian that provided the best agreement with observations favoured 
minimum energy output from the system which, in turn, favours a scenario in which time-averaged heating balances cooling. Below, we 
demonstrate that the minimum energy output Lagrangian actually provides an alternative way of expressing the Gibbs thermodynamic 
potential for this type of system. 

Using equation (\ref{eq:17}), we can express the minimum output energy integral from \cite{pope2011} in terms 
of the GFE,
\begin{eqnarray}\label{eq:19}
E_{\rm out}(t) = \int_{0}^{t} [H(t') + L(t')]\,{\rm d}t' \\ \nonumber = \int_{0}^{t}\bigg\{\frac{\dot{G}(t')}{[1-(v/c_{\rm s})^{2}]} + 2L(t')\bigg\}\,{\rm d}t',
\end{eqnarray}
where $\dot{G}(t)$ is the rate of change of total GFE with respect to time. For a system that is evolving steadily and quasi-statically, 
we assume that $\dot{G} = {\rm constant} < 0$, while $[1-(v/c_{\rm s})^{2}]$ changes sign as a result of i) the work done by gravity
and ii) feedback, such that over long times (and many sign switches) $\int_{0}^{t}\dot{G}/[1-(v/c_{\rm s})^{2}]\,{\rm d}t' \approx 0$. 
In this limit, equation (\ref{eq:19}) reduces to 
\begin{equation}\label{eq:20}
\int_{0}^{t}[H(t') + L(t')]\,{\rm d}t' \approx \int_{0}^{t} 2L(t')\,{\rm d}t',
\end{equation} 
meaning that that time-averaged heating and cooling rates are approximately equal: $\langle H \rangle \approx \langle L \rangle$. 
Thus, the minimum energy output from a galaxy is attained when the time-averaged heating balances cooling such that the system 
evolves quasi-statically towards equilibrium. This finding can explain the seemingly close match between heating and cooling rates
in elliptical galaxies and galaxy clusters inferred from observations \citep[e.g.][]{birzan, best06, nulsen2007, dunn08}.

\cite{pope2011} showed that a direct consequence of the minimum energy output condition is that optimal heating occurs in discrete 
events at periodic intervals, in qualitative agreement with observations \citep[e.g.][]{best05, best07, shab08}. While this is a 
useful result, it is important to remember that it is an idealised heating scenario: in reality AGN duty cycles
will be noisy \citep[e.g.][]{kp06, pope07a, pavpop, gaspari} due to the impact of small-scale processes, but displaying (to varying degrees) the 
features of fuelling processes such as gravitational instability \citep[e.g.][]{sr, king09, popeetal11}, Bondi accretion 
\citep[e.g.][]{allen06,catt07, booth09, mccarthy2010}, or cold feedback \citep[e.g][]{pizz05,soker06} subject to the optimality criterion. 

\subsection{A comment on black hole accretion and entropy}

In any discussion about a system evolving towards thermodynamic equilibrium, it is informative to consider entropy 
changes. For a galaxy, it can be shown that the entropy increase associated with black hole growth far exceeds any entropy 
losses that occur as a result of gas cooling, or gains that result from mass capture by stars.  Specifically, the entropy change 
when a black hole grows by an energy $\dot{m}c^{2}\Delta t$ is \citep[e.g.][]{bek, bard, hawk}
\begin{equation}\label{eq:22}
\Delta S_{\rm BH} = \frac{\dot{m} c^{2}\Delta t}{T_{\rm BH}},
\end{equation}
where $T_{\rm BH}$ is the black hole temperature given by \citep[e.g.][]{hawk}
\begin{equation}\label{eq:23}
T_{\rm BH} = \frac{\hbar c^{3}}{8\pi G_{\rm N} M_{\rm BH}k_{\rm b}}
\end{equation} 
with $\hbar$ being the reduced Planck's constant, $G_{\rm N}$ is Newton's gravitational constant, and $M_{\rm BH}$ is the black hole mass. 
It can be seen from equation (\ref{eq:23}) that the black hole temperature falls as its mass increases, thereby increasing the entropy gain
whenever the black hole accretes material.

In the limit that the black hole dominates the total heating rate, $H$, the power output and black hole 
accretion rate are related by $H = \eta_{\rm c}\eta_{\rm BH}\dot{m}_{\rm BH}c^{2}$. This means an energy $(1-\eta_{\rm BH})\dot{m}_{\rm BH}c^{2} = (1-\eta_{\rm BH})H/(\eta_{\rm c}\eta_{\rm BH})$ 
is captured by the black hole. Therefore, equation (\ref{eq:22}) becomes 
\begin{equation}\label{eq:24}
\Delta S_{\rm BH} =  \frac{(1-\eta_{\rm BH})}{\eta_{\rm c}\eta_{\rm BH}} \frac{H \Delta t}{T_{\rm BH}}.
\end{equation}  
Then, since the temperature of the black hole is much lower than that of the hot atmosphere, accretion onto the black hole causes the entropy of the galaxy to increase while the GFE decreases.

\section{Summary}

The aim of this article has been to better understand the origin of feedback heating processes that are fundamental in governing the evolution of 
gas in massive galaxies and galaxy clusters. To achieve this, the Gibbs thermodynamic potential was calculated for a generalised galaxy model that 
describes the 3 main gravitational influences on the galaxy's gas: i) the dark matter gravitational potential; ii) the stars; iii) the supermassive black hole, 
as shown in figure \ref{fig:gibbs_1}. Differences between global heating and cooling rates result in the movement of gas within the galaxy, re-distributing 
mass and energy between the 3 gravitational components. Using this approach, it was shown that periods of heating and cooling are 
required to ensure that the re-distribution reduces the free energy of the gas overall, such that the system evolves always towards 
thermodynamic equilibrium. The key details and findings are described below.

\begin{enumerate}

\item As the hot atmosphere of a massive galaxy cools by radiating X-rays, it will lose gravitational potential energy while simultaneously being compressed by the overlaying gas. 
In addition, the cool gas will be captured by the stars and the supermassive black hole at the centre of the galaxy. All of these processes will increase 
the energy of the gas, while cooling gas that drops out of the atmosphere will reduce the thermal energy. To ensure that the GFE always decreases, strong feedback 
heating will be favoured whenever the rate at which the gas gains energy (through cooling) exceeds the rate of energy loss that occurs when gas drops out 
of the atmosphere, i.e. when $(v/c_{\rm s})^{2} > 1$. Under these circumstances, the galaxy will be in a heating state, as illustrated by the top right panel of figure \ref{fig:gibbs}.

\item When the gas is heated overall, we expect $v/c_{\rm s}$ to decrease. Therefore, at the point that $(v/c_{\rm s})^{2} < 1$, the system will enter 
a period during which overall cooling is favoured. This is because the rate at which the gas gains energy (through cooling) is less than 
the energy loss that occurs when gas drops out of the atmosphere. Under these conditions, the galaxy will be in a cooling state, as shown in the bottom left panel of figure 
\ref{fig:gibbs}. Furthermore, while the gas cools overall, we expect $v/c_{\rm s}$ to increase. Therefore, at the point that $(v/c_{\rm s})^{2} > 1$, the system will enter 
another period during which overall heating is favoured, and so on. 

\item As a direct consequence of the above, the energetics of the system favour a close balance between average heating and cooling rates, explaining correlations between inferred 
heating and cooling rates in elliptical galaxies and galaxy clusters \citep[e.g.][]{birzan, best06, nulsen2007, dunn08}. More significantly, this means that the galaxy evolves 
quasi-statically towards a state of thermodynamic equilibrium. \emph{It is for this reason that AGN feedback regulates the growth of massive galaxies.}

\item  It has also been shown here that minimising the GFE of gas in a galaxy is consistent with minimising the sum of the heating and cooling outputs from the system, thus providing
a broader theoretical motivation for the optimal AGN heating scenario proposed by \cite{pope2011}.

\item Cycles of AGN activity can be explained by galaxies flipping between the heating and cooling states shown in figure \ref{fig:gibbs}. As described above, this flipping is driven by the variation of 
the parameter $v^{2}$ which is influenced by global heating and cooling rates. By comparison with observations \citep[e.g.][]{best05, best07, shab08}, 
massive galaxies with larger radio AGN duty cycles must spend more time in the heating state than lower mass galaxies. 

\item The same effect can explain the characteristic differences of AGN feedback in cool-core (CC) and non cool-core (NCC) galaxy clusters \citep[e.g.][]{burns, mittal}. According to this model, a 
central AGN located in a NCC clusters must spend more time in the cooling state than a central AGN in a CC cluster. This could occur because of different formation histories meaning that NCC 
clusters have systematically higher gas temperatures relative to the dynamical temperature of the cluster gravitational potential, or because of the effect of thermal conduction \citep[e.g.][]{guo},
or differences in metal content \citep[e.g.][]{dubois11}.

\end{enumerate}

\section{Acknowledgements}
The author would like to thank the referee for helpful comments which improved this work.

\bibliography{database} \bibliographystyle{mn2e}

\label{lastpage}

\end{document}